\documentclass{llncs}

\usepackage{amsmath}
\usepackage{url}
\usepackage{graphicx}
\usepackage{epstopdf}
\usepackage{color}
\usepackage{hyperref}
\usepackage{amsopn}
\usepackage{amssymb}
\usepackage{subfigure}
\usepackage{cite}

\DeclareMathOperator*{\MAX}{MAX}
\DeclareMathOperator*{\MIN}{MIN}
\DeclareMathOperator*{\SOR}{\bigvee}

\DeclareGraphicsExtensions{.pdf,.png}
\graphicspath{{figures/}}

\begin{document}

\title{A Self-Organized Method for Computing the Epidemic Threshold  in  Computer Networks}

\author{
    Franco Bagnoli\thanks{franco.bagnoli@unifi.it}\inst{1}\and
   Emanuele Bellini\thanks{emanuele.bellioni@unifi.it}\inst{2}\and
    Emanuele Massaro\thanks{emanuele.emassaro@epfl.ch}\inst{3}
}
\institute{
    Department  of Physics and Astronomy and CSDC,\\
     University of Florence, via G. Sansone 1, 50019 Sesto Fiorentino, Italy. \\
     Also INFN, Sez. di Firenze.\and
   Department of Information Engineering and CSDC,\\ University of Florence,
    via S. Marta 3, 50139 Firenze, Italy.\and
    HERUS Lab, \'Ecole Polytechnique F\'ed\'erale de Lausanne (EFPL),\\
    GR C1 455 (B\^atiment GR) - Station 2,
    CH-1015 Lausanne,
    Switzerland
}

\maketitle

\begin{abstract}
In many cases,  tainted information in a computer network can spread in a way similar to an epidemics in the human world. On the other had, information processing paths are often redundant, so a single infection occurrence can be easily ``reabsorbed''.  Randomly checking the information with a central server is equivalent to lowering  the infection probability but with a certain cost (for instance processing time), so it is important to quickly  evaluate  the epidemic threshold for each node. We present a method for getting such information without resorting to repeated simulations. As for human epidemics, the local information about the infection level (risk perception) can be an important factor, and we show that our method can be applied to this case, too.  Finally, when the process to be monitored is more complex and includes ``disruptive interference'', one has to use actual simulations, which however can be carried out ``in parallel'' for many possible infection probabilities. 
\end{abstract}



\section{Introduction}
 We deal here with the problem of the spreading of tainted information in an unsupervised computer network, such as algorithmic (high frequency) trading~\cite{HFT},
 
The main competitive advantage (given the same information) is the processing time~\cite{HFT1}, which prevents the possibility of checking the information against a central database. However, in this way a tainted information may quickly spread and ``contaminate'' the whole network, in a way similar to what happens for epidemic  in the human world. We have to consider, however, that in many cases the information is processed in a redundant way, so that the tainted information can actually diffuse only if it is able to survive and spread in the network, much like an infection which has to fight against the defences of hosts. 

Well-known results from the theoretical epidemiology field show that there is a strict relationship between the infection probability $\tau$,  the average number of contacts $\langle k \rangle$ and its variance, i.e.,  $\langle k ^2\rangle$: the critical value $\tau_c$ for the onset of an epidemic is $\tau_c = \frac{\langle k \rangle }{\langle k^2 \rangle}\simeq \langle k \rangle ^{-1}$
for sharp-distributed networks~\cite{risknet}. In many cases however the contact network can be approximated  by a scale-free distribution with diverging variance, for which there is no hope of controlling  epidemics only by reducing the infection probability~\cite{EpidemicSpreadingScaleFreeNetworks,EpidemicsComputerViruses}. 

The influence of  risk perception in epidemic spreading has been studied for human epidemics~\cite{riskperception}, where the knowledge about the diffusion of  disease among neighbours (without knowing who is actually infected)  lowers the effective probability of transmission. For regular, random, Watts-Strogatz small-world and non-assortative scale-free networks with exponent $\gamma >3$ there is always a finite level of precaution parameter for which the epidemic goes extinct~\cite{risknet}. For scale-free networks with $\gamma < 3$ the precaution level depends on the cut-off of the power-law, which at least depends on the finite number of the nodes of the network. 

In humans,  information about the disease may not come  from physical contacts, but rather from  the ``virtual'' social contact networks~\cite{virtual_inf,virtual_inf1,virtual_inf2}.   Clearly, one expects that if these two networks are completely different, the  perception of the risk is of less value than when the two networks coincide. Again, this is a common situation also for automatic trading and computer networks. 

In a computer network, a node  can indeed choose not to accept the processing of an incoming information, but this refusal also has a certain cost.  In other words,  it is sometimes preferable to suspend the information processing than risking the elaboration of false data, according with the cost of such operation. We can model this situation by assuming that the tainted information can propagate with a certain probability, that may depend on the knowledge one has about the infection levels in the network or at least in its neighbourhood. This  infection probability is however also a measure of the cost of processing. In order to lower the infection probability  one may have for instance to contact a central server, lowering al,so the transaction frequency. On the other hand,  information processing paths are often redundant, so a single infection occurrence can be easily ``cured'' by other nodes, assuming that all nodes cooperate, sharing the cost. 

It is therefore vital to quickly assess the epidemic threshold for a given network (that may change in time), with real-time estimates of the infection probability threshold, that may change from node to node. The optimal probability is that just below the epidemic threshold, in which the cost of checking is minimal but the tainted information cannot diffuse and is eliminated in the long time limit by the redundancy of information-processing paths. 

We present here a method (first introduced in Ref~\cite{bagnoli_rech} and extended in Ref.~\cite{multiplex}) that can be applied in such situations.  The proposed method allows to obtain the epidemic threshold in just one run, without having to repeat the simulation with many tentative infection probabilities, looking for the outbreak threshold. This method can be considered an example of self-organized criticality~\cite{SOC}, in which a system automatically discovers the critical value of a parameter. In particular, it is very reminiscent of the   Bak-Sneppen evolutionary model~\cite{BakSneppen}. The proposed method can be directly implemented in computer networks, allowing nodes to exchange also their estimated epidemic threshold.

Epidemic models are characterized by a monotone increasing of the probability of being contaminated with the number of infected neighbours, and this characteristic allows to explicitly obtain the epidemic threshold by the self-organized critical method.  We can consider also other processes, for instance with an ``interference'' among infective agents, and in general  processed based on generic local rules like cellular automata~\cite{EquivalenceCA}. However, in this cases the monotonicity is lost and one has to consider more complex data structures~\cite{fragments}.

\section{The Infection Model}

We consider a set of $N$ nodes $x_i$, with two states: 0 for ``healthy'' and 1 for ``tainted'' (or contaminated). Node $i$ process information coming from other nodes $j$, defined by an adjacency matrix $a_{ij} = 1 (0)$ for connected (disconnected) nodes.  We define the input connectivity of node $i$ as $k_i = \sum_j a_{ij}$. We assume that if a node $i$ is tainted, it can ``infect'' other nodes with a probability $\tau$, that for the moment is fixed. 

Let us start with a simple percolation model: a node $i$ can be infected by each of its $k$ neighbors separately with a probability $\tau$, so that if $s$ of them are infected, the total infection probability $q(s, k_i)$ is 
\[
q(s, k_i) = 1- (1-\tau)^s\simeq s\tau
\]
for small $\tau$.  It is evident that the knowledge of the average number of infected neighbours is a crucial information for deciding whether to process the received information or not.

We assume that after having processed the information,  nodes do not retain any tainted data, so the process is a SIS (Susceptible-Infected-Susceptible) one. We consider a parallel SIS model, which is equivalent to a directed percolation problem where the directed direction is the time. Actually, this is an example of a directed \textit{bond} percolation.

This model is implemented   by computing for each node $i$ and time $t$ the state $x_i(t)$ by taking the OR ($\vee$) of the infection process along each connection, where the single infection event from node $j$ to node $i$ is computed by extracting a random number $r_{ij}$, evenly distributed between 0 and 1, and comparing it with $\tau$, i.e., 
\begin{equation}
x_i(t+1) = \bigvee_{j=j^{(i)}_1, \dots, j^{(i)}_{ k_i}} [\tau> r_{ij}(t)] x_j(t),
\label{eq1}
\end{equation}
where $\bigvee$ represents the OR operator and the multiplication represents the AND. The square bracket represents  the truth function, $[\cdot]=1$ if ``$\cdot$'' is true, and zero otherwise. The quantity $r_{ij}(t)$ is a random number between 0 and 1,  drawn independently for each  triplet $i,j,t$.  

Alternatively one can study the \textit{site} percolation process, where the node $x_i$ first processes all incoming information and then, probabilistically ($\pi$), checks the result. In this case the dynamics is 
\begin{equation}
x_i(t+1) = [\pi> r_{i}(t)] \bigvee_{j=j^{(i)}_1, \dots, j^{(i)}_{ k_i}}  x_j(t).
\label{eqsite}
\end{equation}

In the case of risk perception, we assume that $\tau$ is replaced by a probability $u(s, k_i) $ that a site $i$ with connectivity $k_i$ is infected by any one of its $s$ infected neighbours as 
\begin{equation}\label{f}
  u(s, k_i) = \tau f(s; J),
 \end{equation}
  where $\tau$ is the ``bare'' infection probability and $f(s;J)$ is a monotonic decreasing function of the number of infected neighbours $s$, depending on some parameter $J$. For instance, in Ref.~\cite{riskperception}, the probability $u(s, k_i)$
  was assumed to be 
 \begin{equation}\label{prp}
   u(s, k_i) = \tau\exp\left(-J\frac{s}{k_i}\right), 
\end{equation}

The idea is that the perception of the risk, given by the percentage of infected neighbours and modulated by the factor $J$, effectively lowers the infection probability because the node checks the received information against the central server, paying the delay.

\section{The Self-Organized Percolation Method}
 
The basic idea is that each node $i$ estimates its own minimum value $\tau_i$ of the infection probability, or the maximum value of the the precaution $J_c$ for barely being infected. Iterating this procedure for long time, we find the critical value of the parameters for having the smallest surviving epidemics, for the given choice of  the random numbers $r_i(t)$.

Once chosen, these random numbers behave like a quenched field.  In principle the epidemic threshold is given by the average over the statistical ensemble, i.e., over many repetitions of the processes. However, in many cases the process is self-averaging~\cite{selfaveraging}, i.e., a large enough system gives the same results as the whole statistical ensemble, possibly inducing some small error in the determination of the critical threshold, since in this case the correlation length diverges. 
  
Let us pretend that we are performing several simulations of the bond percolation process in parallel, with different values of $\tau$ but using the same set of random numbers. Due to the ``monotonic'' character of the infection, if, for a given site $i$ and time $t$, the percolation stops for some value of $\tau$, it stops also for all lower values.  We can therefore replace $x_i(t)$ by $[\tau> \tau_i(t)]$ (or $[\pi > \pi_i(t)]$ for the site problem). 
\begin{figure}[t!]
    \begin{center}
        \includegraphics[width=0.9\columnwidth]{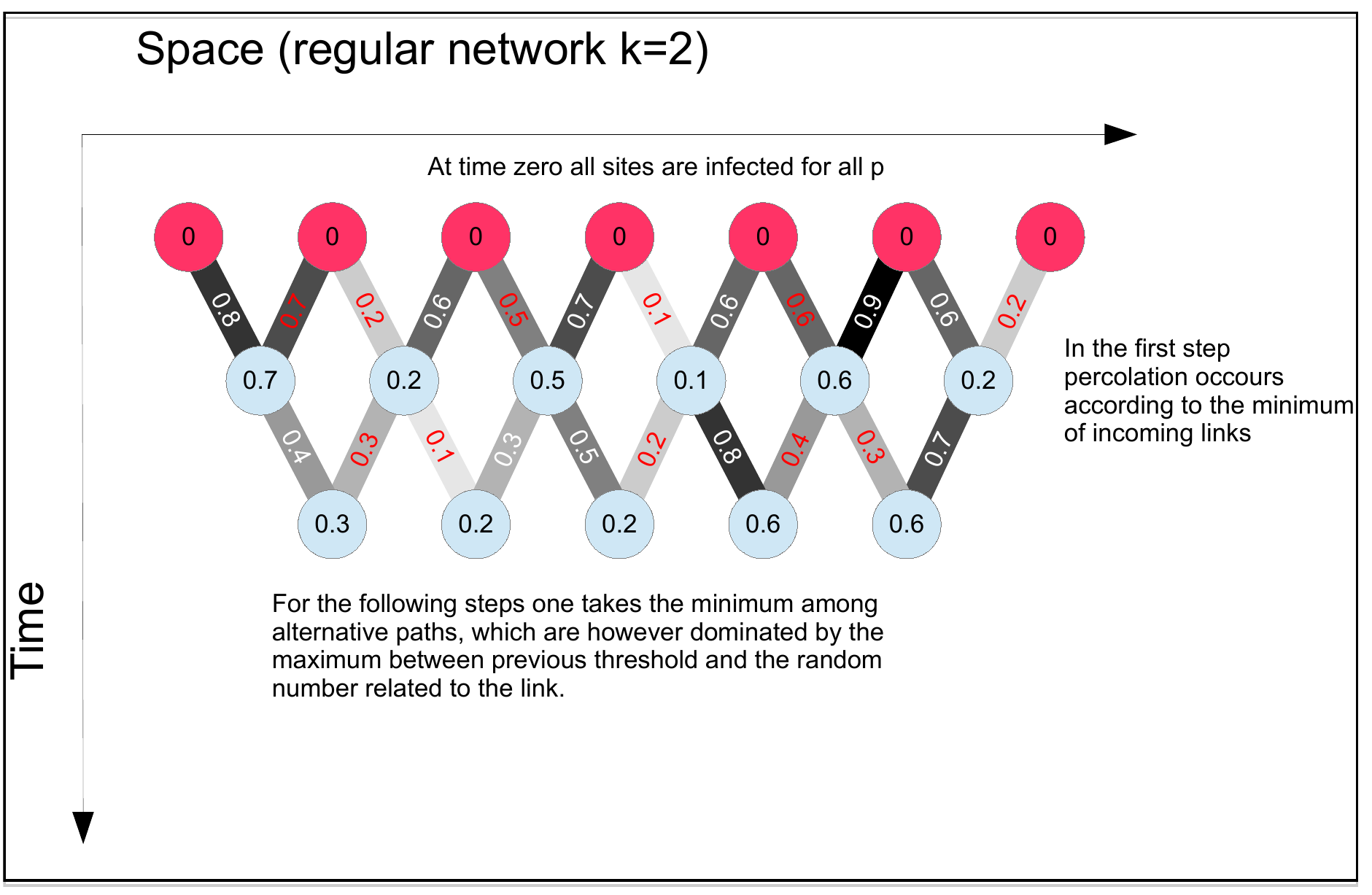} 
    \end{center}
    \caption{\label{fig:schema}  Evolution of the local minimum value of the percolation parameter $p_i$ for a 1D regular network with $k =2$. }
\end{figure}

\begin{figure}[t!]
    \begin{center}
        \includegraphics[width=0.9\columnwidth]{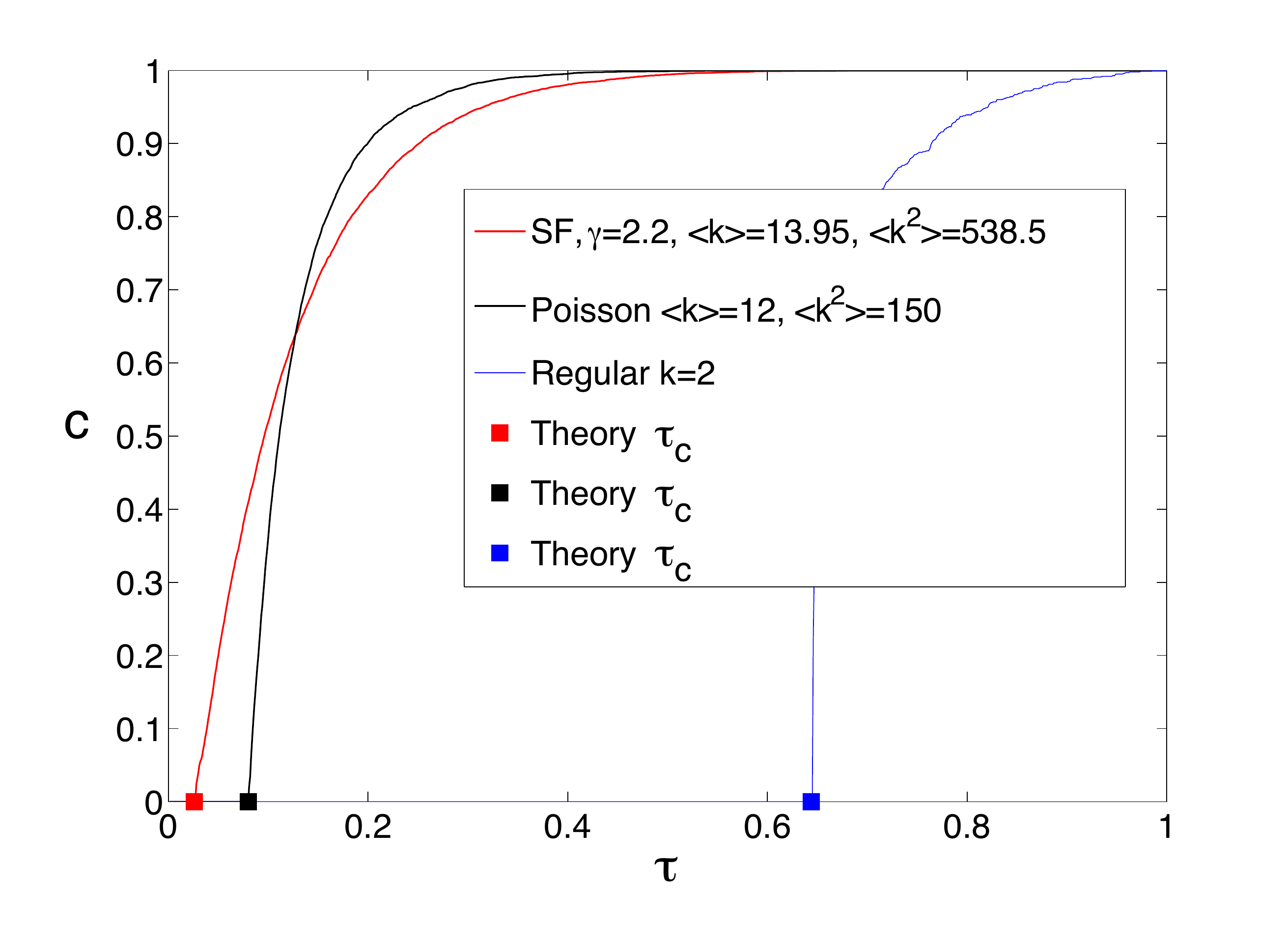}
    \end{center}
    \caption{\label{fig:perc}   Asymptotic  number of infected individuals $c$ versus the bare infection probability $\tau$ for the SIS dynamics for different networks. From left to right, for $c=0$: Scale Free (SF), Random (Poisson), Regular. Here $N=10000$. }
\end{figure}

For the bond  percolation problem, Eq.~\eqref{eq1} becomes:
\begin{equation}
\label{eq2}
  [\tau> \tau_i(t+1)] = \bigvee _{j=j^{(i)}_1, \dots, j^{(i)}_{ k_i}} [\tau>r_{ij}(t) ] [\tau>\tau_j(t)].
\end{equation}
Since $[\tau>a ] [\tau>b]$ is equal to $[\tau>\max(a,b)]$ and $[\tau>a] \vee [\tau>b]$ is equal to $[\tau>\min(a,b)]$, Eq.~\eqref{eq2} becomes:
\begin{equation}
\label{eq3}
  [\tau>\tau_i(t+1)] =\left[\tau> \left(\MIN_{j=j^{(i)}_1, \dots, j^{(i)}_{ k_i}} \max\bigl(r_{ij}(t),\tau_j(t)\bigr)\right)\right],
\end{equation}
and  we get the desired equation for the $\tau_i$'s
\begin{equation}
\label{p}
   \tau_i(t+1) = \MIN_{j=j^{(i)}_1, \dots, j^{(i)}_{ k_i}} \max\bigr(r_{ij}(t),\tau_j(t)\bigl).
\end{equation}
Let assume that at time $t=0$ all sites are infected, so that $x_i(0)=1\; \forall \tau$. We can therefore write $\tau_i(0)=0$. We can  iterate Eq.~\eqref{p} and get the asymptotic distribution of $\tau_i$. The minimum  of this distribution gives the critical  value $\tau_c$ for which there is at least one percolating cluster with at least one ``infected'' site  at large times, i.e., there is an epidemic spreading in the whole system.  This procedure is illustrated in Fig.~\ref{fig:schema} for a regular lattice in 1 dimension and $k=2$. 

For site  percolation, the equivalent equation is 
\begin{equation}
\label{pi}
\pi_i(t+1) =  \max\Bigr(r_{i}(t),\MIN_{j=j^{(i)}_1, \dots, j^{(i)}_{ k_i}}\pi_j(t)\Bigl).
\end{equation}

We investigated the  SIS dynamics over regular, Poisson and scale-free networks as shown in Fig.~\ref{fig:perc}. In particular we evaluated the critical epidemic threshold values $\tau_c$ for which there is at least one percolating clusters with at least one infected nodes (points marked "Theory $\tau_c$ in  Fig.~\ref{fig:perc}).

Considering a regular lattice with  connectivity degree $k=2$, we found $\tau_c\simeq 0.6447$ which is compatible with  the results of the bond percolation transition in the Domany-Kinzel model~\cite{DK}.

In the case of random networks with Poisson degree distributions the critical epidemic threshold is $\tau_c= \langle k \rangle /  \langle k^2 \rangle  \simeq \langle k \rangle ^{-1} $ if the distribution is sharp~\cite{epidemic2}. Indeed, for a Poisson network with $\langle k \rangle = 12$ the self-organized percolation method gives $\tau_c \simeq 0.08 \simeq 1/12$. 

For a scale-free  network with $\langle k \rangle = 13.95$ and $\langle k^2 \rangle = 538.5$ we get from simulations $\tau_c \simeq 0.026$, in agreement with the expected value. 

\begin{figure}[t!]
    \begin{center}
        \includegraphics[width=0.9\columnwidth]{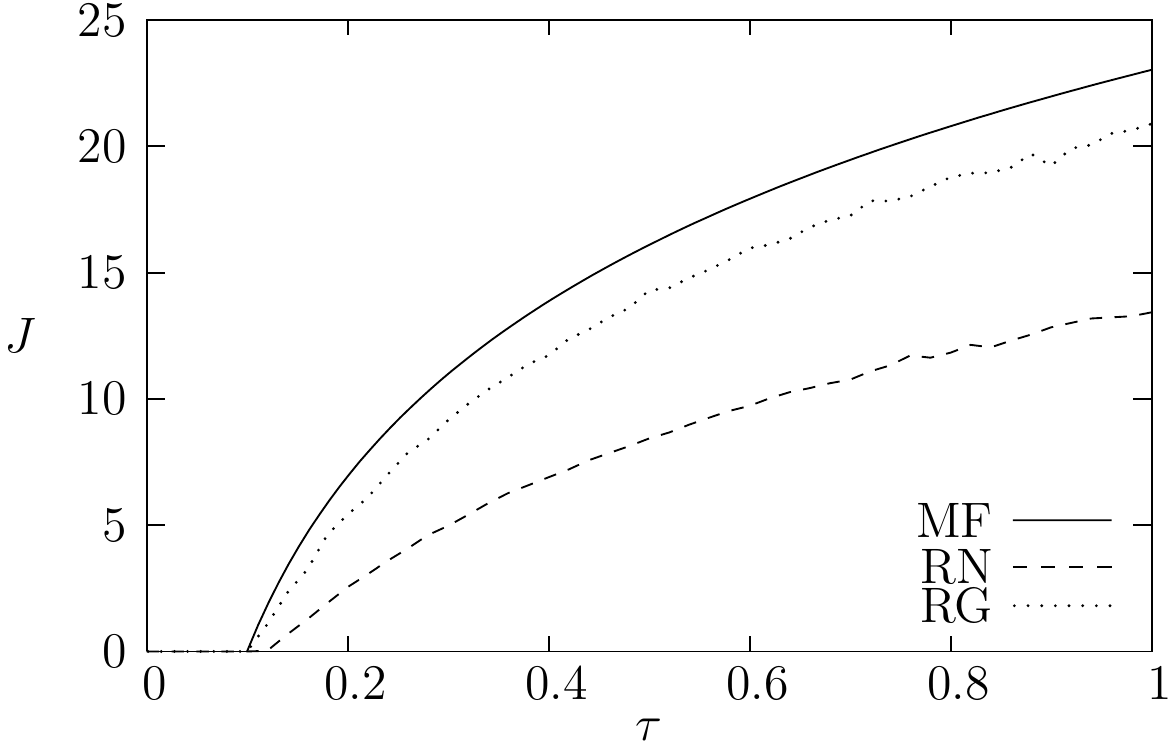}
    \end{center}
    \caption{\label{fig:Jc-tau} Critical level $J_c$ for which the infection is stopped, for networks with fixed or peaked connectivity $k = 10$ and $N = 1000$ in the mean-field (MF), regular (RN) and random (RG) case.} 
\end{figure}

Now, let us apply the method to a more difficult problem, for which the percolation probability depends on the fraction of infected sites in the neighbourhood (risk perception), es expressed by Eq.~\ref{f}.
In this case we want to find the extremal value of the parameter $J$ for which there is no spreading of the infection at large times. 

Again, we can replace $x_i(t)$ by $[u>u_i(t)]$ and invert the relation $u_i(t)= \tau f(s;J) $ so that that at the end one gets an equation for the $x_i(t)$ like $[J \lessgtr J_i(t)]$ which can be iterated. 

Let us consider for illustration the case of Eq.~\eqref{prp}. The quantity $[u>r]=[\tau \exp(-J s/k) > r]$ is equivalent to $[J   < - ( k/s) \ln(r/\tau)]$. Therefore Eq.~\eqref{eq2} is replaced by 

\begin{equation}
\label{pprp}
 [J < J_i(t+1)] 
= \SOR_{j=j^{(i)}_1, \dots, j^{(i)}_{ k_i}} \left[ J < - \frac{k_i}{s_i} \ln\left(\frac{r_{ij}(t)}{\tau}\right)\right] [J < J_j(t)] 
\end{equation}
where
\begin{equation}
\label{s}
    s_i\equiv s_i(J) = \sum_{j=j^{(i)}_1, \dots, j^{(i)}_{ k_i}} x_{j} = \sum_{j=j^{(i)}_1, \dots, j^{(i)}_{ k_i}}       [J_j(t)\ge J]. 
\end{equation}
So
\begin{equation} 
 \label{ppprp}
 [J < J_i(t+1)] =
\SOR_{j=j^{(i)}_1, \dots, j^{(i)}_{ k_i}} \left[ J < - \frac{k_i}{s_i(J_j(t))} \ln\left(\frac{r_{ij}(t)}{\tau}\right)\right] [J < J_j(t)] 
 \end{equation}
and therefore 
\begin{equation}
\label{rp}
J_i(t+1) =   \MAX_{j=j^{(i)}_1, \dots, j^{(i)}_{ k_i}} \min\left( - \frac{k_i}{s_i(J_j(t))} \ln\left(\frac{r_{ij}(t)}{\tau}\right), J_j(t)\right).
\end{equation}

Analogously to the previous case, the critical value of $J_c$ is obtained by taking the maximum value of the $J_i(t)$ for some large (but finite) value of $t$.

\begin{figure}[t!]
    \begin{center}
            \includegraphics[width=0.9\columnwidth]{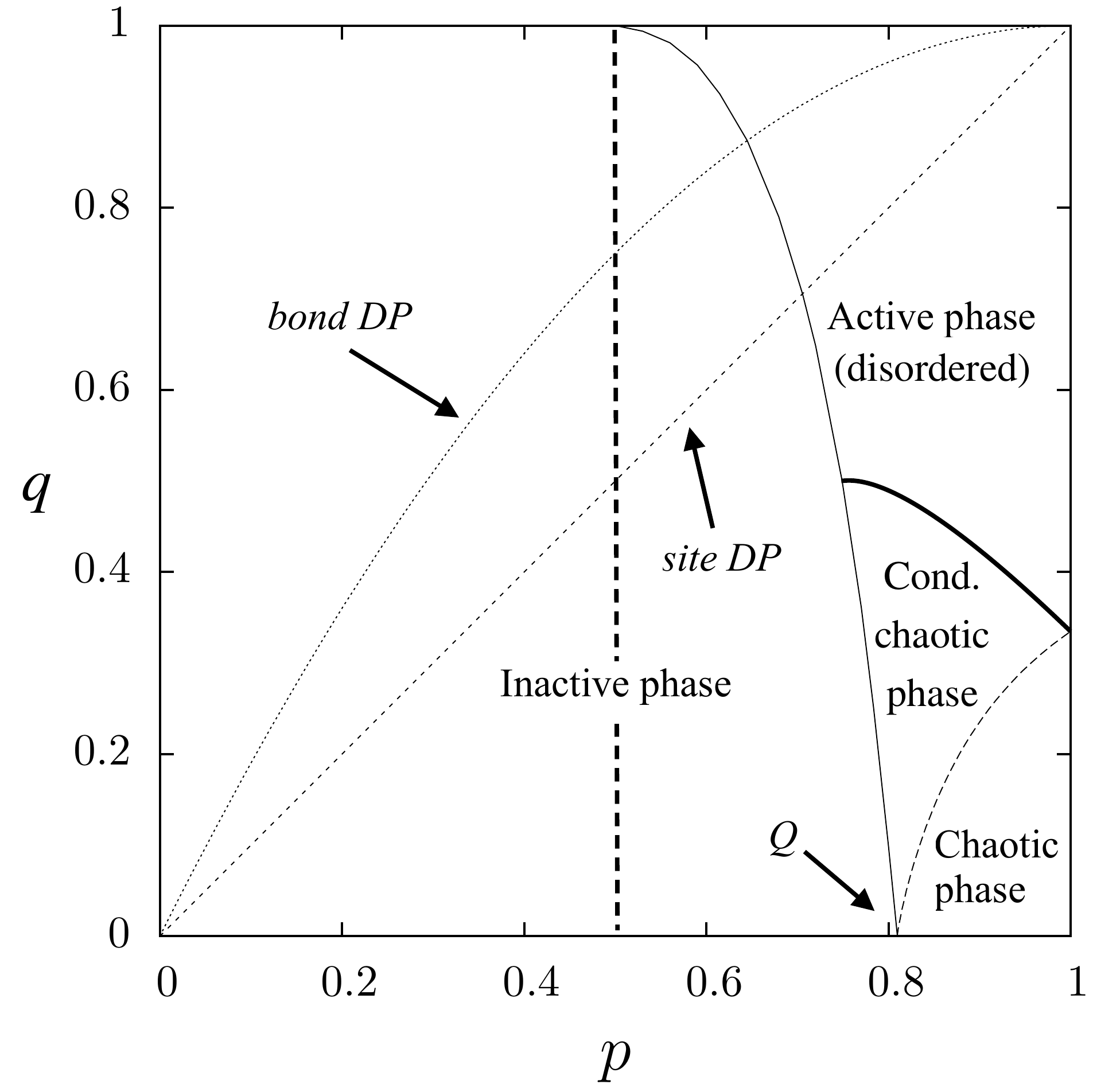}
    \end{center}
    \caption{\label{fig:DK} The phase diagram of the Domany-Kinzel cellular automaton model. In the quiescent phase only the state with 0 infected sites is stable. In the active phase the state 0 is unstable and the average number of infected sites is larger than 0. In this phase the long-time evolution only depends on the initial condition, so it may be defined disordered. In the conditionally chaotic and chaotic phase the evolution depends on the initial condition and therefore varies when the configuration is varied (therefore ``chaotic''). In the chaotic phase this dependence occurs for all implementations, in the conditionally chaotic one only for some particular computational scheme.} 
\end{figure}

\begin{figure}[t]
    \begin{center}
        \begin{tabular}{cc}
            \includegraphics[width=0.48\columnwidth]{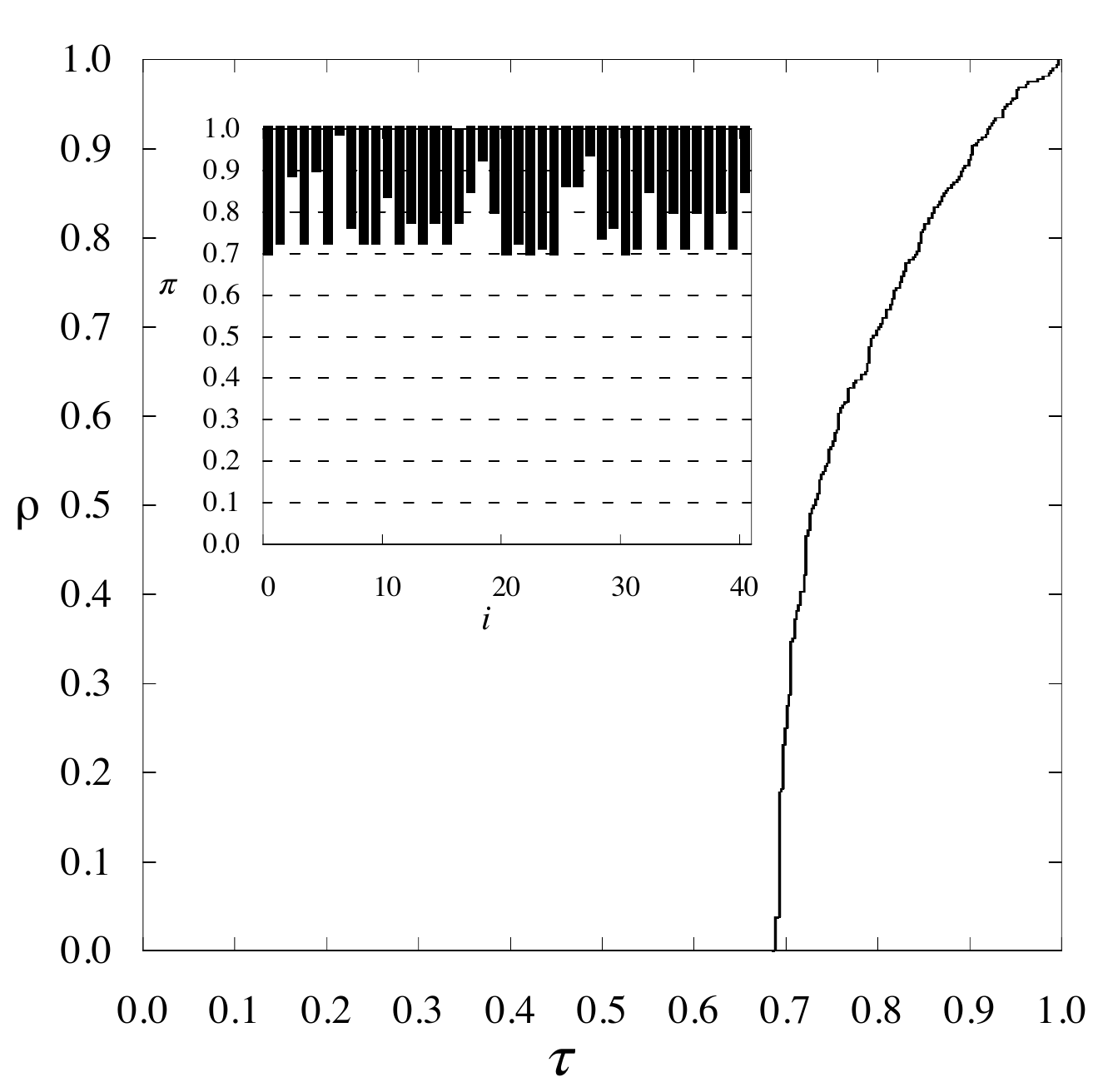} &
            \includegraphics[width=0.49\columnwidth]{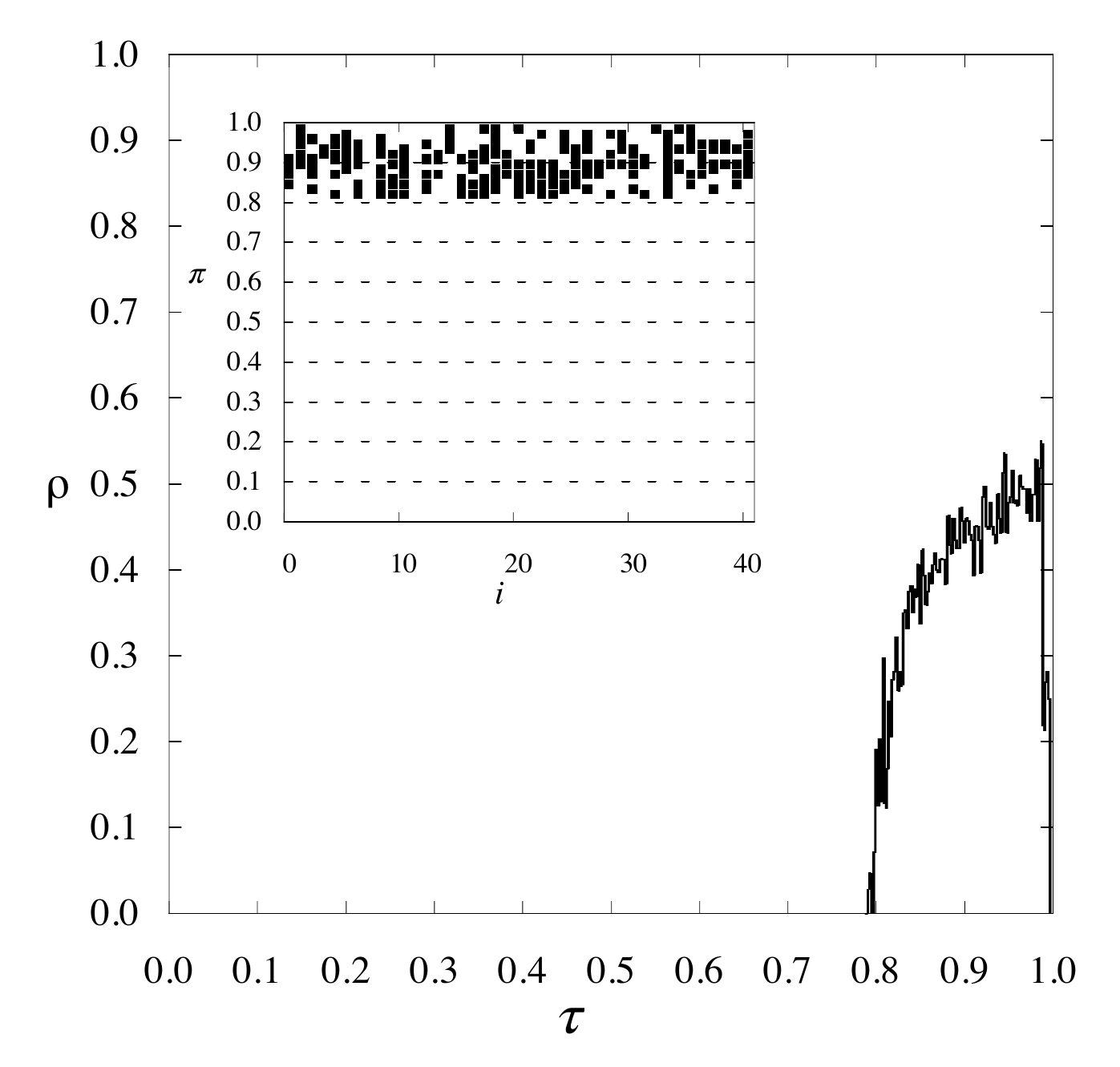}
        \end{tabular}
    \end{center}
    \caption{\label{fig:fragment}  Left: the average density ($\rho$) as a function of the bare infection probability $\tau$ and in the inset the distribution of the values of $\pi_i$ for which the infection reaches site $i$ (black bar) for the site percolation problem in a regular network with $k=2$ (DK model, $p=q$). Right:  the average density ($\rho$) as a function of the bare infection probability $\tau$ and in the inset the distribution of the values of $\pi_i$ for which the infection reaches site $i$ (black bar) for the a ``nonlinear'' (XOR) percolation problem in a regular network with $k=2$ (DK model $q=0$).} 
\end{figure}

\begin{figure}[t!]
    \begin{center}
            \includegraphics[width=0.9\columnwidth]{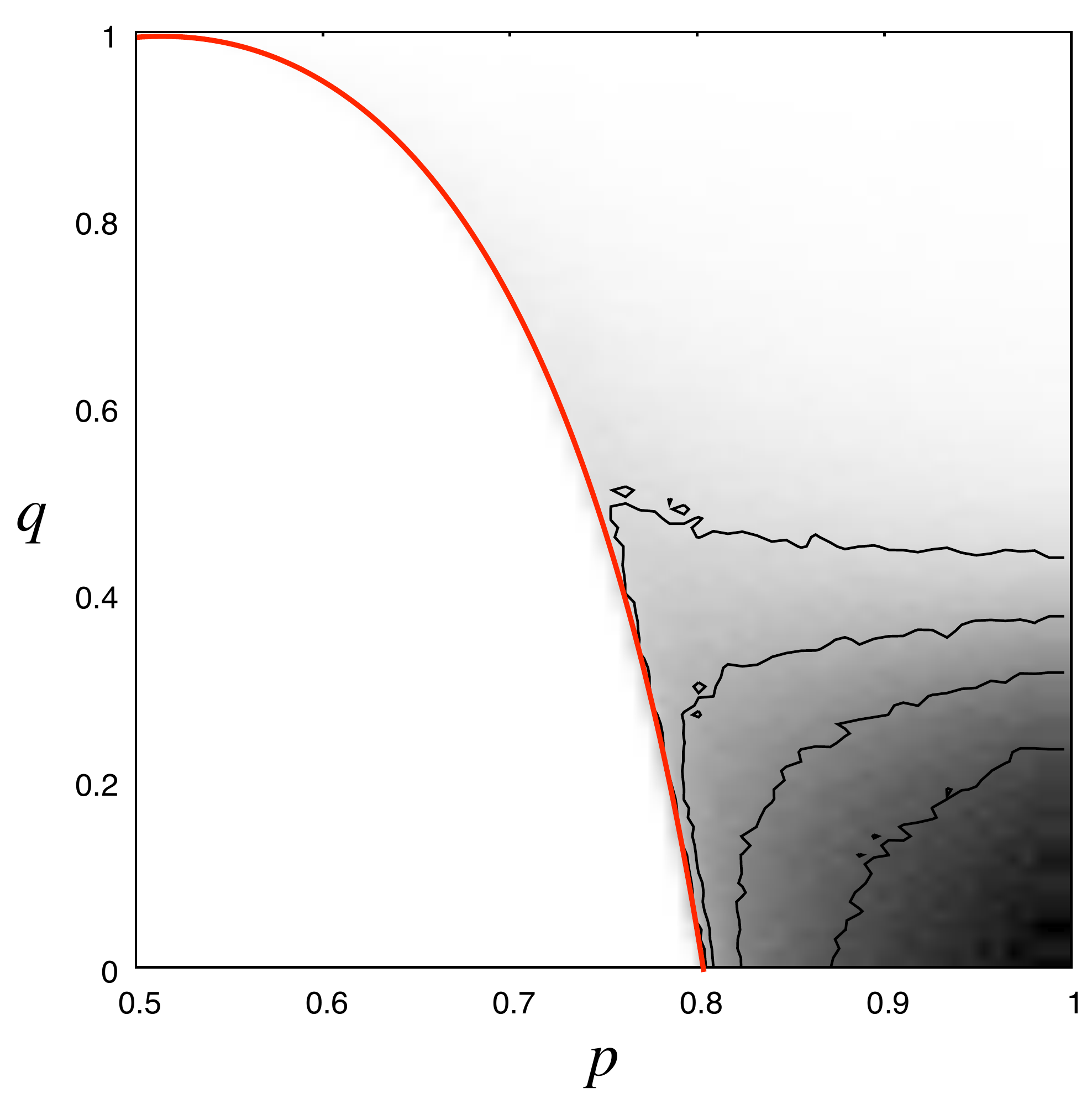}
    \end{center}
    \caption{\label{fig:dens} Regions in the DK model where the infection probability   shows negative variation (non-monotonicity). The implementation scheme coincides with that giving the conditionally chaotic phase in Fig.~\ref{fig:DK}. The contour curves are for  the  density of negative variation equal to  0.10, 0.15, 0.20, 0.25. Computation for 10,000 time steps,  $N=10,000$,  $q$ fixed in steps of $0.01$ and $p$ sampled in 64 equally-spaced points from $0.5$ to 1.} 
\end{figure}

The results are quite interesting compared with the simple SIS dynamics, for which there is always an epidemic threshold (Fig.~\ref{fig:perc}). By inserting the risk perception it is possible to stop the epidemic for every value of the bare infection probability $\tau$ up to $\tau=1$, for networks with finite variance.  Let us consider for instance the case of random networks with $\langle k\rangle=10$; for which for the simple infection process we found  a critical value  $\tau_c=0.165$. As shown in  Fig.~\ref{fig:Jc-tau}, beyond  this value of $\tau_c$ the epidemics can still be stopped if all agents adopt a sufficiently high precaution level $J$. The same consideration can be done also for the other scenarios.

However, as reported in Ref.~\cite{riskperception}, for some scale-free networks even the perception of the risk is not able to stop the epidemics, and one has to resort to more specialized techniques, like using special precautions for hubs, which is what is usually done also in the computer world.

\section{Non Monotonic Infection Probability}

The self-organized method for finding the epidemic threshold relies on the monotonicity of the infection probability $f$ with respect to the considered parameter, Eq~\eqref{f}. 

However, not all processes fulfils this requirement. In particular, if there is a ``disruptive interference'' among possible spreaders that diminishes the infection probability, it may happen that a larger number of infected neighbours actually slows down the epidemics. 

In order to illustrate this problem, let us consider the Domany-Kinzel model~\cite{DK}.  It is probably the simplest model on a regular one-dimensional lattice with nearest-neighbours interactions still presenting an interesting phase diagram~\ref{fig:DK}. The DK model is a totalistic cellular automaton with $k=2$ inputs, so it is defined by 3 transition probabilities $\tau(1|n)$ which is the probability that a site will be infected in the following time step if $n$ of its neighbours are infected, $n=0,1,2$. Since the appearance of new infected individuals in a healthy population is a rare event, we set $\tau(1|0)=0$. The other two parameters are $p=\tau(1|1)$ and $q=\tau(1|2)$. 

This model generalizes the bond and site percolation problems, as shown in Fig.~\ref{fig:DK}. Above the line marked ``bond DP'' there is a synergistic infectious effect: the probability of being infected by two contaminated  neighbours is higher that the ``superposition'' of the two separate events. Below the line marked ``site DP'' there is an interference effect, and the probability of being infected by two simultaneous contaminated neighbours is less than the probability of being infected by just one of them. 

The self-organized method works above the site DP line, as illustrated in Fig.~\ref{fig:fragment} for $q=p$ and $q=0$. In the insets, the asymptotic distribution of the infection for sites and for all values of the parameter $p$ is shown. One can see that the corresponding segments are compact for the site percolation problem $q=p$, white they are fragmented for $q=0$. This means that in the first case one can simply iterate the computation for the lower end of the segment, which is the essence of the self-organized method. 

For the rest of the phase diagram, one can resort to a parallel computation for many values of the parameters, simply by coding the possible statuses using multi-bit technique, as described in Ref.~\cite{fragments}. In Fig.~\ref{fig:fragment} the result of such computation keeping fixed $q$ and sampling $p$ using 64 bits (indicated by $p_j$, $j=1,\dots,64$) is reported. The quantity shown is the number of ``holes'' in the segments, i.e., the number of times for which is a given site one has infection for a certain value $p_j$ while the site is not infected for $p_{j+1}$.  One can see that the region for which the infection probability shows such negative changes is concentrated around the corner $p=1$, $q=0$, i.e., where the interference effect is larger. By comparison with Fig.~\ref{fig:DK}, is seems that this region coincides with the chaotic one, i.e., the region in which the evolution of the system depends also on the initial conditions, and not only on the choice of the random numbers~\cite{ObjectiveDefinitionDamageSpreading,BagnoliOnDamage}.

\section{Conclusions}

We  investigated the problem of epidemic spreading of tainted data on computer networks, exploiting a self-organized method, that automatically gives the percolation threshold in just one simulation. 

We showed that this method can be extended by considering the knowledge of the local infection level, and that this element may allow to halt an infection even for large ``bare'' infection probabilities. The method can be extended also to the case in which the knowledge about the infection comes from sourced partially different from the ones that actually communicate the ``disease'', provided that this difference is  not too large.  

Finally, we  considered the case of  more complex processes, including  ``disruptive interference'' among spreaders. In this case the probability of being infected for a given site at an asymptotic time is not monotonous with the control parameters , and our self-organized method cannot be used. In these cases one can however exploit the self-averaging character of the problem, and carry out parallel simulations using multi-bit coding and just one or two  random numbers per site.  For the Domany-Kinzel cellular automaton, the  region for which the self-organized method is not applicable seems to coincide with the ``chaotic'' one, for which the evolution is not only given by the choice of random numbers, but is still dependent on the initial state. 

In the future, we shall work to develop a security protocol based on such  scheme and test it  on more realistic computer networks and processes.

\bibliography{biblio}{}
\bibliographystyle{unsrt}

\end{document}